\documentclass[graybox]{svmult}

\usepackage{mathptmx}       
\usepackage{helvet}         
\usepackage{courier}        
\usepackage{type1cm}        
\usepackage{makeidx}         
\usepackage{graphicx}        
\usepackage{multicol}        
\usepackage[bottom]{footmisc}
\usepackage{enumerate}       

\usepackage[linesnumbered, ruled]{algorithm2e} 

\makeindex             

\begin{document}

\title*{ComPass: Proximity Aware  Common Passphrase Agreement Protocol for Wi-Fi devices Using Physical Layer Security }
\titlerunning{ComPass}
\author{Khan Reaz and Gerhard Wunder}
\authorrunning{K. Reaz and G. Wunder}
\institute{Khan Reaz, Gerhard Wunder \at Institute of Computer Science, Freie Universit\"at Berlin, Germany, \email{kahn.reaz@ieee.org}}
%
%
\maketitle

\abstract*{Secure and scalable device provisioning is a notorious challenge in Wi-Fi. WPA2/WPA3 solutions take user interaction and a strong passphrase for granted. However, the often weak passphrases are subject to  guessing attacks. Notably, there has been a significant rise of cyberattacks on Wi-Fi home or small office networks during the COVID-19 pandemic. This paper addresses the device provisioning problem in Wi-Fi (personal mode) and proposes \textit{ComPass} protocol to supplement WPA2/WPA3. \textit{ComPass} replaces the pre-installed or user-selected passphrases with automatically generated ones. For this, \textit{ComPass} employs Physical Layer Security  and extracts credentials from common random physical layer parameters between devices. Two major features make \textit{ComPass} unique and superior compared to previous  proposals: First, it employs  phase information (rather than amplitude or signal strength) to generate the passphrase so that it is robust, scaleable, and impossible to guess. Our analysis showed that \textit{ComPass} generated passphrases has 3 times more entropy than human generated passphrases (113-bits vs. 34-bits). Second, \textit{ComPass} selects  parameters such that two devices bind only within a certain proximity ($\leq$3m), hence providing practically useful in-build PLS-based authentication. \textit{ComPass} is available as a kernel module or as full firmware.}

\abstract{Secure and scalable device provisioning is a notorious challenge in Wi-Fi. WPA2/WPA3 solutions take user interaction and a strong passphrase for granted. However, the often weak passphrases are subject to  guessing attacks. Notably, there has been a significant rise of cyberattacks on Wi-Fi home or small office networks during the COVID-19 pandemic. This paper addresses the device provisioning problem in Wi-Fi (personal mode) and proposes \textit{ComPass} protocol to supplement WPA2/WPA3. \textit{ComPass} replaces the pre-installed or user-selected passphrases with automatically generated ones. For this, \textit{ComPass} employs Physical Layer Security  and extracts credentials from common random physical layer parameters between devices. Two major features make \textit{ComPass} unique and superior compared to previous  proposals: First, it employs  phase information (rather than amplitude or signal strength) to generate the passphrase so that it is robust, scaleable, and impossible to guess. Our analysis showed that \textit{ComPass} generated passphrases have 3 times more entropy than human generated passphrases (113-bits vs. 34-bits). Second, \textit{ComPass} selects  parameters such that two devices bind only within a certain proximity ($\leq$3m), hence providing practically useful in-build PLS-based authentication. \textit{ComPass} is available as a kernel module or as full firmware.}

\section{Introduction}
\label{sec:intro}
Connectivity is the key to the world of business, entertainment, education, and government services. While cellular dominates mobility use-case, 802.11 a.k.a Wi-Fi is the single most widely used technology to access the internet when it comes to streaming movies to the smart TV at home, making a video conference call at the workplace, or merely sharing vacation photos from a hotel room or a café. In  recent years, consumers have also embraced Wi-Fi for connecting new types of peripherals as part of their  daily life such as Amazon Alexa powered Echo devices or Google Connected Home or Apple Home accessories.

Quite recently, the world has faced  COVID-19 pandemic. Due to the lockdown, people relied  on home Wi-Fi more than ever  to work remotely. Interpol reported   an alarming rate of cyberattacks  during the pandemic months~\cite{interpol-covid}. The increased number of remote-working has made an adversary  more interested in the radio part of the communication  since it is more straightforward to capture packets over the air.

The Wi-Fi Alliance has developed several security protocols over the last decades to secure Wi-Fi communication. Nevertheless, none of the protocols provides full-proof and future-proof security. Recently,  a significant flaw, popularly known as KRACK-attack  was discovered, and it  heavily affected all platforms~\cite{KRACK}. To ease the provisioning of credentials, especially for resource-constrained devices, Wi-Fi Alliance had developed \textit{Wi-Fi Protected Setup (WPS)} protocol. It  gives consumers an easier option to set up a secure  Wi-Fi connection by pushing a button (PBC mode), or entering a PIN, or via NFC interface \cite{wps}. However, WPS has a long-standing weak security, known as WPS PIN recovery~\cite{wpsCert}.

In an effort to strengthen the security of Wi-Fi, WPA3 has been recently announced (last release v3.0 on December 2020)  \cite{alliance2020wpa3}. The new standard mandates a higher cryptographic standard (192-bit key for enterprise mode, although  128-bit for personal mode). It replaces the \textit{Pre-Shared Key (PSK)} exchange with \textit{Simultaneous Authentication of Equals (SAE)} and introduces \textit{Forward Secrecy}. However, a passphrase is still used. The newly introduced   \textit{Wi-Fi Easy Connect} \cite{EasyConnect} replaces all previous methods of WPS with a Public Key Cryptography (PKC) based provisioning mechanism. In  \textit{Wi-Fi Easy Connect}, a network owner is presumed to have a primary device \textit{(Configurator)} with a rich user interface (e.g., a smartphone or tablet with a camera) that runs the \textit{Device Provisioning Protocol (DPP)}. Here, all \textit{Enrollees} have electronic or printed QR codes or human-readable strings. The \textit{Configurator} scans the code (the user can also manually type in the human-readable strings) to provision the \textit{Enrollee} with credentials. \emph{DPP} relies on QR code scanning, which is not at all feasible for a large number of devices (Think of a premise to be monitored with a Wi-Fi IP camera; then all the cameras have to be scanned and connected to the network). The Wi-Fi Alliance has released another  supporting protocol, called \textit{Enhanced Open}~\cite{RFC8110}. It is an adapted version of the \textit{Opportunistic Wireless Encryption (OWE)}~\cite{OWESpecs} protocol that aims  to mitigate attacks on open un-encrypted wireless networks. Here, the client (STA) and the access point (AP) generate pairwise secret by performing  Diffie-Hellman (DH) key exchange during the 4-way handshake procedure.  OWE is based on PKC, and PKC is  threatened by the uprising of quantum computers~\cite{keyfactor}. It is to be noted that the exponent size used in the DH must be selected such that it has at least double the entropy of the entire crypto-system, i.e., if we use a group whose strength is 128-bits, we must use more than 256-bits of randomness in the exponent used in the DH calculation~\cite{rfc3526}. This brings to the required DH key-size of 4200 bits at its best strength estimation~\cite{rfc3526}.  The large key-size is a massive burden for the  IoT ecosystem~\cite{keyfactor}.

In recent years several works have been done to generate the secret key using PHY-layer properties based on Shannon's~\cite{shannon1949communication},  Wyner's~\cite{wyner1975wiretap}, and Maurer's~\cite{maurer1993secret} seminal information-theoretic security concept. In  the \emph{Physical Layer Security (PLS)} approach, the inherent reciprocity property of the wireless channel and its varying nature (i.e. randomness) is used to agree on a key between two legitimate transceivers. Enthusiasm among researchers gave a significant rise  towards developing key generation algorithms on this principle. Most of the existing works are based on \emph{Amplitude} or \emph{Received Signal Strength (RSS)}~\cite{xi2016instant,thai2018secret,zenger2014novel, zenger2015security}. It is because   \emph{Amplitude} and \emph{RSS} show reciprocity without much effort and hence can be easily reconciled to generate a symmetric key. On the other hand, the slightest displacement of the transceivers cause the \emph{Phase} to  vary significantly. Xi et al. proposed the Dancing Signal (TDS) scheme in~\cite{xi2016instant}.  It requires devices to be within 5cm which is very impractical since most of the cases APs are wall-mounted or hidden to keep away unwanted hardware access. In TDS, keys are generated from the local entropy source instead of the randomness of the wireless channel. Their evaluation showed a good performance since the implementation is done on a traditional computer. This will not be the case for resource-constrained IoT devices which are known to have low entropy~\cite{keyfactor}. From the literature, it is well established that \emph{Amplitude} or \emph{Received Signal Strength (RSS)} based existing methods are slow, need iterative communications, authenticated channel, and a large number of samples to generate a good quality  key.

We propose \textit{ComPass} to tackle the  challenges  mentioned above. It is  a new proximity aware common passphrase agreement protocol for deployable  Wi-Fi network consisting of all classes of Wi-Fi devices (hence, some devices may have no camera or keypad). Our PLS based proposed method uses \emph{Phase} information  of the wireless channel and its varying nature (i.e., randomness) to agree on a passphrase between two legitimate transceivers. With the \emph{ComPass} generated passphrase, it is possible to generate 128/192/256-bit (or higher) key with high entropy at a minimum communication overhead. Our intention is not to replace the well known WPA2/WPA3; instead, supplement it with the new automated passphrase generation protocol.

The paper is organized as follows. A brief introduction to Wi-Fi channel measurement is given in Sec.\ref{sec:preliminaries}. We presented the end-to-end steps  of the \textit{ComPass} protocol in Sec.\ref{sec:compass}. and its security analysis in Sec.\ref{sec:security_analysis}.  Sec.\ref{sec:implementation} describes the implementation details. Our concluding remark is given in Sec.\ref{sec:conclusion}.

\section{Preliminaries}
\label{sec:preliminaries}

We revisit some of the core technologies of the Wi-Fi PHY, specifically \emph{Beamforming}. It utilizes the knowledge (i.e., Channel State Information (CSI)) of the MIMO channel to improve the receiver's throughput significantly.

In the complex baseband MIMO channel model, a vector $x_{k} = [x_1, x_2, ... x_{N_{T_x}}]^T$ is transmitted in subcarrier $k$  using  OFDM scheme. The received vector $y_k = [y_1, y_2,... y_{N_{R_x}}]^T$ is then modeled as:
\begin{equation}
    y_k = H_kx_k + Z
\end{equation}
 $H_k$ is the channel response matrix of dimensions $N_{R_x} \times N _{T_x}$ where $N_{R_x}$ is the maximum number of receiving antenna, $N_{T_x}$ is the maximum number of transmitting antenna. $H_k$ is expressed in  complex number to represent the attenuation (i.e amplitude $(\|H_k\|)$ and the phase shift $(\angle H_{k}))$ for each subcarrier $k$. $Z$ is  the additive white Gaussian noise. The  CSI is expressed as a multidimensional matrix taking  $ H = [N_{R_x}]  [N_{T_x}] [N_{S_c}]$ form. $N_{S_c}$ is the number of used data subcarriers~\cite{xie2016atheros,zhu2018pi}. Depending on the Wi-Fi chip, protocol version, bandwidth and channel estimation method, the size of this matrix will vary. For example,  a $3\times 3$ MIMO device with a Qualcomm Atheros Wi-Fi chip operating on IEEE 802.11n 5GHz band with a $ BW $=20/40 MHz would report CSI as a $ [3][3][56/114] $ matrix. We refer to the IEEE standard~\cite{80211_2016} for the detailed explanation of the IEEE 802.11 PHY procedure.

\section{ComPass Protocol}
\label{sec:compass}

Let us define  entities of  the \emph{ComPass} protocol. \emph{Access Point} is kept hidden (to reduce Evil Twin attacks) and it has an \emph{Authenticator} with a rich user interface. The \emph{Enrollee} is a device with  limited interface (it can have a rich user interface too). Before initiating the protocol, devices are brought within  proximity ($\leq3$m). Summary of the protocol steps are as follows 

\begin{enumerate}[(1)]
    \item With a button press  or after booting, the \emph{Enrollee} broadcasts its name-id with random nonce in Wi-Fi infrastructure mode.  Power button or existing WPS button can be re-programmed for this purpose.

    \item \emph{Authenticator} verifies and confirms the \emph{Enrollee} from an app or  from the system's Wi-Fi setting.
    \begin{enumerate}[a.]

        \item \emph{Authenticator} and \emph{Enrollee} perform procedures as mentioned in the following sections (\ref{subsection:ssc} to \ref{subsection:mapping}) to generate a common $passphrase_1$. Once connected, the \emph{Authenticator} sends (SSID + AP-MAC)  to the \emph{Enrollee}. Subsequently, it sends Enrollee’s MAC+ $passphrase_1$ to the \emph{Access Point}. This communication  is already encrypted since the \emph{Authenticator} has joined the network beforehand.

        \item \emph{Enrollee} switch to Wi-Fi Client mode after receiving (SSID + AP-MAC) from the \emph{Authenticator}. It sends \emph{Association} request to the \emph{Access Point} appending hashed $passphrase_1$.

        \item \emph{Access Point} verifies the request by comparing hashed $passphrase_1$. If successful, it initiates procedures as described in Sec.~(\ref{subsection:ssc} to \ref{subsection:mapping}).
    \end{enumerate}
    \item \emph{Access Point} and \emph{Enrollee} generates $passphrase_2$ in the similar way.

    \item If successful, \emph{Access Point} allows \emph{Enrollee} to connect and it notifies \emph{Authenticator}, else \emph{Enrollee} returns to step (2)b. Finally, \emph{Authenticator} and \emph{Access Point} delete the $passphrase_1$.
\end{enumerate}

\emph{Authenticator} and \emph{Enrollee} refer to STA and the \emph{Access Point} as AP. We assume that the \emph{Authenticator} joins the \emph{Access Point} securely either by existing WPA2/WPA3 method or by generating their  common passphrase according to the  procedures mentioned in Sec.~(\ref{subsection:ssc} to \ref{subsection:mapping}). New devices can only be joined  through the \emph{Authenticator(s)}. In the following subsections we present the intermediate steps of the protocol and  algorithms.

\subsection{Synchronized CSI Collection }
\label{subsection:ssc}

In the last few years, several toolchains have been developed by researchers to extract CSI from commercial off-the-shelf (COTS) devices. Among them, the Intel CSI Tool (ICT) by Halperin et al. \cite{halperin2011tool} and the Atheros CSI Tool (ACT) by Xie et al. ~\cite{xie2016atheros} are  widely used. The recent release of \emph{nexmon} CSI extraction tool~\cite{gringoli2019free} has opened the door for extracting CSI from Broadcom and Cypress chipsets. Although there are some differences between these toolchains,  they all report CSI to the firmware's user-space in a similar fashion. Hence, \emph{ComPass} remains compatible with all of them.  In this paper, we worked with ACT to implement ComPass on devices. We have    patched some of the bugs that we found in ACT. For example, previously, the driver reported CSI for all packets, including the acknowledgment packets (ACK). It caused one device to have more CSI data than the other. The ACT supports up to 3 RF chains, but the SoC firmware sometimes  use the \emph{Link Adaptation} technique (especially in LOS scenarios) to turn off some antennas. Also, the time stamp associated with the reported CSI was  according to  each device's local clock. It   caused misalignment for our intended use of the CSI to generate a common passphrase on both devices.

One of the first challenges of PLS method is to ensure that the collected channel measurements are coming from the packets that are exchanged within the channel's coherence time. This is to make sure that the collected channel measurements on both sides hold reciprocity property.
To mitigate the unwanted effects on  CSI, we employed a  \emph{Synchronous CSI Collection (SCC)}  procedure between the devices to ensures that they have a common time stamp (up to a certain accuracy) and only CSI from the correct probing packets  are logged. At first, STA aligns its local clock with AP by utilizing the Linux built-in library.  AP instructs STA to start  exchanging a fixed $ N $ number of dummy packets after waiting for $t_d$ seconds. Once CSI for an incoming packet is reported, it checks for $R_x \times T_x$ combination. If $R_x \neq T_x$, CSI value is dropped. After collecting CSI for $N$ packets, the protocol moves to the \emph{Parameter Extraction} step.

\subsection{Parameter Extraction }
\label{subsection:parameterExtraction}

A vast body of literature on channel-based key generation, specifically those who implemented their schemes on COTS hardware relied only on the  \emph{Amplitude/RSS} part of a signal; only a very few considered to work with the \emph{Phase} part~\cite{thai2018secret,wang2012cooperative}.  However, the \emph{Amplitude} fluctuation of the signal is very low in proximity and in an static environment~\cite{autokey}.  An active adversary can generate a synthetic channel amplitude profile to mimic the intended transceiver. Conversely, \emph{Phase} varies significantly  in an indoor environment while respecting the reciprocity property~\cite{wu2015phaseu}. Thus, it is nearly impossible for an adversary to generate a synthetic phase profile. In this  paper, we investigate the \emph{Phase} part of the channel frequency response.

It is to be noted that the CSI reported by the Wi-Fi SoC driver contains the channel's cumulative  frequency response and the device's inner circuitry response as it goes through amplification, down-conversion, packet detection phase. All this additional processing contaminate the true channel response as verified by previous works~\cite{zhu2018pi,wu2015phaseu,kotaru2015spotfi,vasisht2016decimeter,xi2016instant, jung2005time}. Hence, the collected CSI needs sanitizing to remove  unwanted effects.

According to Zhu et al.~\cite{zhu2018pi}, the measured phase $\phi_{k} = \angle H(f_{k})$ can be decomposed as:

\begin{equation}
\small
    \phi_{k} = atan\Bigg(\epsilon_{g} \cdot \frac{sin(2\pi \cdot f_{s} \cdot k \cdot \zeta + \epsilon_{\theta})}{cos(2\pi \cdot f_{s} \cdot k \cdot \zeta)} \Bigg) - 2\pi \cdot f_{s} \cdot k \cdot \lambda + \beta
\label{eq:decomposition}
\end{equation}

where gain mismatch and phase mismatch is denoted by $\epsilon_{g}$, and $\epsilon_{\theta}$ respectively. Unknown timing offset and phase offset error is indicated by $\zeta$ and $\beta$. $\lambda$ sums up the  delay caused by time-of-flight (TOF), packet detection delay (PDD) and  sampling frequency offset (SFO). Note that AWGN is omitted since it would cancel out when comparing phases of the measured CSI from two nodes.

We adapted the decomposition method of~\cite{zhu2018pi} to extract the relevant  parameter from the measured CSI phase. We have  studied the characteristics of these five parameters through several measurement campaigns performed at various locations at the Freie Universität Berlin and other private apartments that included LOS and NLOS scenarios. Our key findings are: (i) The almost sigmoidal-shaped arcus tangent function of the Eq. \ref{eq:decomposition} strongly conforms in the LOS scenario and fails in the NLOS scenario. (ii) Cumulative delay parameter, $\lambda$ is almost constant, which is expected because TOF, PDD, SFO remains static for a low mobility environment. Conveniently, $\lambda$ could be useful to filter out CSI for a packet that arrived later than the channel coherence time $T_{c}$. (iii) $\epsilon_{g}$ and $\epsilon_{\theta}$ are the only useful parameter with good statistical properties.

This revelation of our analysis encouraged us to extract $\epsilon_{g}$ and $\epsilon_{\theta}$ from the collected CSI and proceed to the next steps. Taking the Eq. \ref{eq:decomposition} as a reference decomposition model, we estimate the default value for each of the five parameters from the ideal \emph{arctan} function: $\epsilon_{g}= 0.512, \zeta = -0.02812, \epsilon_{\theta}=-0.006355, \lambda=-0.02762, \beta=0.1326$. Then we perform a non-linear least square curve-fitting operation to estimate the parameters.

\begin{algorithm}
\caption{\small Related algorithms for ComPass protocol}
\label{alg:parameter}
\DontPrintSemicolon

\SetAlgoLined
\SetNoFillComment

\tcc{Step 1: Delay Aware Parameter Extractor (DAPPER)}
    \SetKwInOut{Input}{Input}
    \SetKwInOut{Output}{Output}

    \Input{CSI-phase:  $ \phi_{k} $ of $ N $ packets }
    \Output{Parameter:  $ \epsilon_{g}$}

    \For{$i \leftarrow 1$ to N}
    {
    Extract  $ \epsilon_{g}, \zeta, \epsilon_{\theta}, \lambda, \beta $ by non-linear least-square curve fitting on Eq.\ref{eq:decomposition}

    \eIf  {$\lambda_{i} \gg \lambda_{0}$}{
      Drop the associated $ \epsilon_{g}$

      }{
        Continue
        }
    }

\tcc{Step 2: Moving Window based quantizer (MOW)}

    \Input{Channel parameter $ \epsilon_{g}$}
    \Output{Quantized bits $Q_A $ for STA, $Q_B$ for AP}

    \textbf{Get} the mean $RTT$ from exchanged $N$ packets

    \textbf{Calculate} window size, $ w  \leftarrow \lceil {mean(RTT)}\rceil $

    \For{$j \leftarrow 1$ to N/w}
    {
        \For{$i \leftarrow 1$ to $ w $}
        {
            \textbf{Calculate} mean, $\bar{w} \leftarrow \sum_{i} ^{w}  \epsilon_{g} $\\\
            $Q_{A_{i}/B_{i}} = \left\{ \begin{array}{rcl}1 & \mbox{for}& \epsilon_{g_i} \geq \bar{w} \\ 0 & \mbox{for} & otherwise \\
            \end{array}\right.$\\
        }
    }

\tcc{Step 3: Generate Secure Sketch at STA}

       \SetKwInOut{Input}{Input}
        \SetKwInOut{Output}{Output}

        \Input{$Q_{A}$, random strings $R_{1}, R_{2} $}
        \Output{$S_{S}$}

        Generate $R_{1}, R_{2}$:  where $len(R_{1} = R_{2}) = len (Q_{A})$\\

        Multiply $R_1$ and $R_2$ to get  $M_{R_{1}R_{2}}$ \\

        Use BCH  to generate  code: $C \leftarrow BCH_{enc}(M_{R_{1}R_{2}})$\\

        Generate syndrome of the code: $syn(C)$\\

        \Return $S_{S}(Q_A, M_{R_{1}R_{2}}, syn(C)) $

\tcc{Step 4: Recovery at AP}

        \SetKwInOut{Input}{Input}
        \SetKwInOut{Output}{Output}

        \Input{$Q_{B},S_{S}$}
        \Output{ $Q_{A}$}

        Generate $K(Q_{B},S_{S})$

        Decode with BCH: $N \leftarrow BCH_{dec}(K)$

        \Return{$Q_{B}(S_{S}, N) \equiv Q_{A}$}

\end{algorithm}

Before we implemented \emph{ComPass} on our COTS setup, we used a simulation tool for the next steps by quantizing both $\epsilon_{g}$ and $\epsilon_{\theta}$. Our analysis showed that  $\epsilon_{g}$ gives a slightly better result. Henceforth, $ \sum_{i = 1} ^{N}  \epsilon_{g}$ is the parameter  from the measured CSI-phase that we will use in the following steps. The  \emph{Delay Aware Parameter Extractor (DAPPER)} algorithm is described in the Step 1 of Algorithm~\ref{alg:parameter}. AP and STA perform \emph{DAPPER} independently.

\subsection{Parameter Quantization }
\label{subsection:quantization}


 Existing \emph{lossy} and \emph{lossless} (as categorized by Zenger et al. in~\cite{zenger2015security}) quantization schemes in the literature tend to overlook the fact that the underlying reciprocity would be broken if the \emph{guard-interval} for converting measured complex-valued vectors to  bit-string is calculated based on the whole CSI data set. Keeping this fact in mind, we opted in for an adaptive moving window based quantizer (MOW) (Step 2 of Algorithm~\ref{alg:parameter}). It is a lossless scheme and produces bit-string at 1 Bit/sample. The resulted scheme overcomes the well-known problem of burst 0's and 1's (i.e., $000\ldots 0$, $111 \ldots 1$).

In an one-hop wireless environment,  \emph{Round-Trip-Time (RTT)} can be a useful metric to roughly estimate the effective channel coherence time $(T_{c})$ instead of using the Clarke's mathematical reference  model~\cite{rappaport2001wireless}: $T_c = \sqrt{\frac{9}{16\pi ({f_{m}})^2}}$,~($f_m$ is the Doppler spread). \emph{RTT} is readily available for each packet, and it takes into account various factors including propagation delay, clock offset, processing delay, motions of objects in the environment. We get the mean \emph{RTT} value for the exchanged packets to set the window size $w$ for the MOW quantizer, which is then rounded up according to the IEEE 745 standard respecting the half-to-even rule. The minimum is $w = 3$ since it needs at least 3 packets to successfully  calculate the distance for two nodes  (with asynchronous clocks). Then starting from the most significant bit, we take $w$ element from $\epsilon_g$ and find the mean $\bar{w}$ of that window. We convert each element of the $w$ to  $1/0$  such that $Q_{A_{i}/B_{i}} = \left\{ \begin{array}{rcl}
1 & \mbox{for}
& \epsilon_{g_i} \geq \bar{w} \\ 0 & \mbox{for} & otherwise \\
\end{array}\right.
$.
After that, it moves to the next window and continues until the last element. If the last window   has fewer elements than $w$,  it will be filled by $0$. This process will construct quantized bit strings $Q_A$ for STA and $Q_B$ for the AP.

\subsection{ Reconciliation }
\label{subsection:reconciliation}

Reconciliation shares the common properties of error-correction.
The quantized bits on AP and STA are not necessarily the same; thus they cannot be used \textit{as is}. In~\cite{dodis2008fuzzy}, Dodis et al. presented a new primitive: \emph{Secure Sketch (SS)}. We employ  \textit{SS} as the  reconciliation protocol for its  notable advantages over others~\cite{dodis2008fuzzy}. It allows reconciling one party's quantized bits with the other at minimum leakage. We chose a binary Bose–Chaudhuri–Hocquenghem (BCH) code based construction for \emph{SS}, referred to as PinSketch~\cite{harmon2008implementation}. It  is  the most efficient, flexible, and  linear  over $GF(2)$. One can overcome the computation time by choosing an efficient decoding algorithm for the BCH~\cite{dodis2008fuzzy}. We designed the algorithm  in a bottom-up approach using the available BCH library in the Linux kernel~\cite{bchcodec}.

\subsubsection{Secure Sketch:}
\emph{SS}  generates public information $X$ about its input $a$ that can be used to reproduce $a$ from its correlated version $a'$, where $a \in \mathcal{M} $ and the metric space $\mathcal{M}$ has a distance function $\delta$. It is a randomized procedure involving $\mathsf{Sketch}$ and $\mathsf{Recover}$ such that for input $a \in \mathcal{M}$, $\mathsf{Sketch}$ produces a string $s \in \{0, 1\}$.  The $\mathsf{Recover}$ procedure,  $\mathsf{Recover} (a',\mathsf{Sketch}(a)) = a $ works when $  \delta (a, a') \leq t$, $t$ is the number of error. It uses  random bit strings  to mask original information from an adversary.

\subsubsection{Construction Procedure:}

At this point, STA and AP has quantized bit strings  $Q_{A}$, and $Q_{B}$ respectively which are similar but not same. Our goal is to reconcile $Q_{B}$ with $Q_{A}$ at minimum leakage. We start designing the algorithm by choosing the Galois field order $m$. In our case $m= 7$ for generating a 128-bit key; which makes the maximum BCH codeword size $n = 127 \leftarrow (2^m-1)$. Details of the BCH algorithm is out of scope of this paper, hence, we refer to the original works~\cite{bose1960class},~\cite{hocquenghem1959codes} and its modified version for \emph{SS} in~\cite{harmon2008implementation}. With the optimum error-correcting capability set as $t=9$ bits, we create blocks each with 56 bits resulting 3 blocks. Because of the size of $n$, the last block has  padding bits. Then each block is treated independently to produce secure sketch according to the Step 3 of Algorithm~\ref{alg:parameter} and  concatenated:  $S_{S} \leftarrow S_{s_{1}} \| S_{s_{2}} \| S_{s_{3}} \|$

STA sent $S_{S}$ to AP as the helper string (note that $S_S$ does not expose the quantized bits $Q_A$). AP  performs $\mathsf{Recovery}$ operation according to the Step 4 of Algorithm~\ref{alg:parameter} to find the mismatch in $Q_{B}$ and correct them. Usually, in a BCH decoder, error locator root-finding is done by Chien search~\cite{chien1964cyclic}. However, in our implementation,  we used the technique of~\cite{biswas2009efficient} for its better performance. It consists of factoring the error locator polynomial using the Berlekamp Trace algorithm down to degree 4. After that, the low degree polynomial solving technique of~\cite{zinoviev1996solution} is used. Fianally, AP and STA possess the same bit string, resulting in $Q_B \equiv Q_A$.

\subsection{Mapping bits to passphrase}
\label{subsection:mapping}
We map each 8-bit (starting with MSB) of the $Q_A/Q_B$ according to the widely adopted 8-bit \emph{Unicode} (UTF-8) (i.e., total 256 characters) encompassing the whole alphabet set of a passphrase (lowercase, uppercase, numerals, and symbols). Since there are some control and non-latin characters within the UTF-8 table, we changed U+0000 -- U+0020 $\rightarrow$ uppercase HEX,  and U+0080 -- U+00FF $\rightarrow$ lowercase HEX. U+0021 -- U+007E remains unchanged. This way, the generated passphrase complies with password policies such as lower and uppercase letters, digits and symbols (converted HEX are treated as regular AlphaNumeric). Finally, the resulted passphrase is treated as per the IEEE 802.11 standard's recommended passphrase to PSK mapping, as defined in IETF RFC 2898 section 5.2~\cite{kaliski2000rfc2898}.

\section{Security Analysis:}
\label{sec:security_analysis}

\begin{figure}[htbp]
\centering
    \includegraphics[scale=.25]{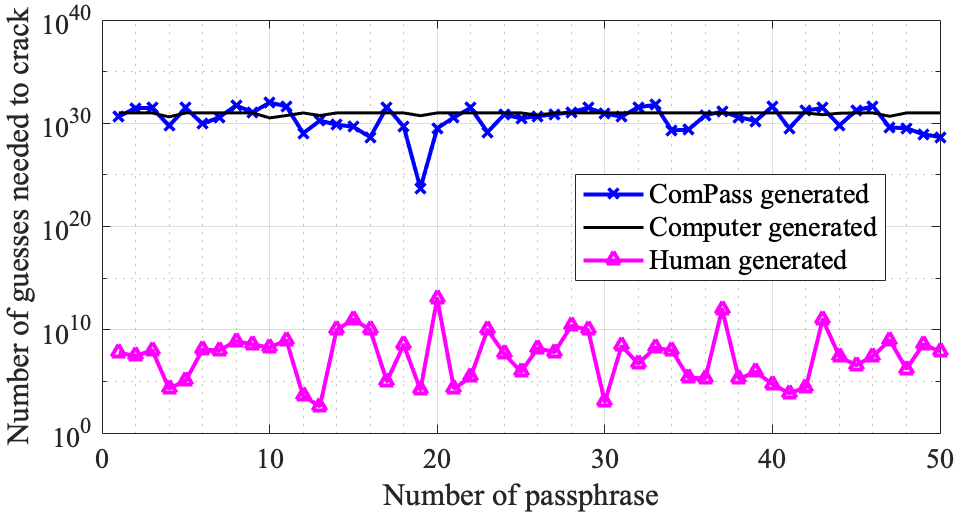}
    \caption{Guess analysis of passphrases generated by \emph{ComPass}, machine, and human.}
    \label{fig:guess_analysis}
    \vspace{-1em}
\end{figure}

\begin{figure}[htbp]
\centering
        \includegraphics[ scale=0.45]{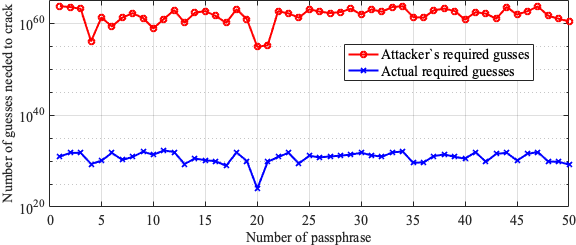}
        \caption{ Guess analysis of passphrases between an attacker and the STA }
        \label{fig:guess_attacker}
        \vspace{-1em}
\end{figure}

\begin{figure}[htbp]
\centering
    \includegraphics[scale=.5]{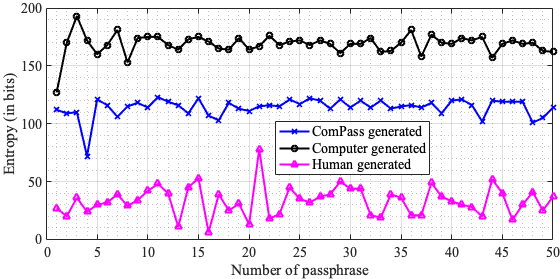}
    \caption{Available entropy (in bits) from  \emph{ComPass}, machine, and human generated passphrases. }
    \label{fig:entropy_analysis}
    \vspace{-1em}
\end{figure}

We used two well-known password quality estimators to evaluate \emph{ComPass}  generated passphrase. Microsoft's \texttt{zxcvbn} toolkit~\cite{wheeler2016zxcvbn} is used to calculate the number of minimum attempts needed to guess (crack) a password using brute-force. \texttt{zxcvbn}'s algorithm finds \emph{token, reversed, sequence, repeat, keyboard, date, and brute force} pattern to estimate strength (as shown in Fig.~\ref{fig:guess_analysis}, and Fig.~\ref{fig:guess_attacker}). \texttt{KeePass}-- recommended by the German Federal Office for Information Security (BSI-E-CS001/003 1.5), and audited~\cite{keepass} by the European Commission's Free and Open Source Software Auditing (EU-FOSSA 1) project is used to calculate the available entropy (as shown in Fig.~\ref{fig:entropy_analysis}). We assume that the information leakage due to reconciliation is negligible and at most $ t  log_2  (n+1) $ (as mentioned in Theorem 6.3 of \cite{dodis2008fuzzy}). Notably, an upper bound  is given by 56 bits in our case. Since it is a different metric than the password strength, we leave its evaluation for future work. We have collected 50 Wi-Fi passphrases from various Cafes, Hotels, and users, which we label as the human-generated passphrase. Then we use an Apple Macbook Pro (with dedicated crpto processor) to generate another set of 50 passphrases using OSX Keychain's Password Assistant tool. Finally, we compare these two sets with 50 \emph{ComPass} generated passphrases for AP (Bob) and STA (Alice).

In Fig.~\ref{fig:guess_analysis}, it is shown that the human-generated passphrases would need less than $ 10^{15} $ attempts, whereas machine-generated passphrases almost always need $ 10^{31} $ attempts to crack it using brute-force. \emph{ComPass}-generated passphrases  went up as high as $ 10^{32} $ and never below $ 10^{24} $ guesses.

To evaluate an attacker's (Eve) performance, we put Eve very close ($ \leq $ wavelength/2) to the Alice and generate 50 passphrases for Bob-Eve. Although Eve is closely located to Alice, the \emph{Phase}  part of her channel profile is very different from Alice's (as also observed by Wu et al. in ~\cite{wu2015phaseu}). Whereas, \emph{Amplitude} and \emph{Signal Strength} of the two is very similar. For this very reason, we chose to work with the \emph{Phase} (as we have explained earlier).

Now, we compare Eve's passphrases with Alice's. We append the actual (Alice-Bob's) passphrase with   Eve's and Alice's to mimic the fact that Eve has partial knowledge of the channel profile. Appending Alice-Bob's  passphrase to Alice does not make a difference since \emph{repeat} is recognized by \texttt{zxcvbn}, and \texttt{KeePass}. Eve's channel profile will be the product of Alice-Bob's channel profile ($H_{AB}$) and Bob-Eve's channel profile ($H_{EA}$). Eve cannot separate it without a  noiseless secondary channel. Notice from Fig.~\ref{fig:guess_attacker} that the  chances of Eve to guess the valid passphrase would be very low as the number of guesses is drastically high even though Eve's channel profile consists of Alice-Bob's channel profile.

We positioned  the STA at different distances; (1, 3, 5, 10, 15)m apart from the AP in various  indoor environments to verify the proximity aspect of \emph{ComPass},  . We observed that the reconciliation scheme (in Sec.~\ref{subsection:reconciliation}) fails when the distance is greater than 3m. It happens due to the multi-path effect that causes the reciprocity phenomenon to break, and thus there left almost negligible common randomness in the channel-phase profile to generate a common passphrase.

Using \texttt{KeePass} entropy analysis tool, we show that on an average the human-generated passphrases have  34-bit entropy,  \emph{ComPass}-generated ones have 113-bit, and machine-generated passphrases have 168-bit entropy (Fig.\ref{fig:entropy_analysis}). Thus \textit{ComPass} generated passphrases have nearly 3 times more entropy than a typical human generated passphrase.

\subsection{Outlook on Privacy Amplification}

In a conventional channel-based key generation methods, a final step called \emph{Privacy Amplification} is performed to cover the lost entropy during the \emph{Reconciliation}. We forgo this additional step in our current implementation of \emph{ComPass} protocol in favor of a \emph{Secure Sketch} based reconciliation protocol, which inherently provides security against leakage. In our future work, we aim to incorporate the \emph{KECCAK} algorithm based  NIST SHA-3 family hash functions for this purpose~\cite{dworkin2015sha3}. After this step, we hope to see that the notches in the curve of the guess analysis of  \emph{ComPass} generated passphrase is reduced.

\section{Implementation}
\label{sec:implementation}

Our  demo setup involves implementing the   algorithms on COTS hardware. We chose very ordinary and widely available TP-Link N750 routers (v1.5, v1.6), and Android device (8.0+) playing the role of  AP and STA.  We were operating our devices in 802.11n and chose channel number 40 (on 5GHz) with BW = 20MHz. Our patched version of ACT uses the upgraded \emph{ath10k} driver instead of  \emph{ath9k}. All of the devices were equipped with $3 \times 3$ antenna, and \emph{Modulation and Coding Scheme (MCS)}- index 16 is set to enable transmission with all 3 antennas. For the non-linear least-squares fitting, we have used the \emph{least-square-cpp} library by \cite{leastsquare}. We enabled the bidirectional channel estimation option, where two devices (regardless of their role) exchange sounding \emph{Physcial Layer Protocol Data Unit (PPDU)}. The receiving STA computes an estimate of the MIMO channel matrix  $H_k$ for each subcarrier $k$ and for each RF chain. While it is possible to extract key-bits from all the available 9 antenna combination, we have implemented one of the nine paths for the demo. We put our devices in various co-working rooms of Freie Universität Berlin campus and private apartments resembling typical indoor environments to perform measurements and protocol tests.

\section{Conclusion}
\label{sec:conclusion}
We presented \emph{ComPass}, a PLS inspired common passphrase agreement protocol for all classes of Wi-Fi devices governed by proximity ($\leq3$m). It forgoes the necessity of memory friendly short password generation by an user and the dependency on PKC. We showed that the \emph{ComPass} generated passphrase has increased  the number of guesses required to crack it using brute force or dictionary attack compared to a typical human-generated  passphrase, and it has increased the available entropy 3 times (113-bits vs. 34-bits). \emph{ComPass} has been implemented on COTS hardware running the latest OpenWrt. The compiled module is 143kb in size, and  can be installed on existing devices using \emph{opkg} package manager or as a full firmware replacement.


%

\bibliographystyle{styles/spmpsci}
\bibliography{ComPass.bib}

\begin{thebibliography}{10}
\providecommand{\url}[1]{{#1}}
\providecommand{\urlprefix}{URL }
\expandafter\ifx\csname urlstyle\endcsname\relax
  \providecommand{\doi}[1]{DOI~\discretionary{}{}{}#1}\else
  \providecommand{\doi}{DOI~\discretionary{}{}{}\begingroup
  \urlstyle{rm}\Url}\fi

\bibitem{biswas2009efficient}
Biswas, B., Herbert, V.: {Efficient Root Finding of Polynomials over Fields of
  Characteristic 2.}
\newblock Avilable at \url{https://hal.archives-ouvertes.fr/hal-00626997/}
  (2009)

\bibitem{bose1960class}
Bose, R.C., Ray-Chaudhuri, D.K.: On a class of error correcting binary group
  codes.
\newblock Information and control \textbf{3}(1), 68--79 (1960)

\bibitem{chien1964cyclic}
Chien, R.: {Cyclic decoding procedures for Bose-Chaudhuri-Hocquenghem codes}.
\newblock IEEE Transactions on information theory \textbf{10}(4) (1964)

\bibitem{bchcodec}
Djelic, I., Borgerding, M.: {User BCH (Bose-Chaudhuri-Hocquenghem)
  encode/decode library based on BCH module from linux kernel}.
\newblock Available at \url{https://github.com/mborgerding/bch\_codec} (2015)

\bibitem{dodis2008fuzzy}
Dodis, Y., Ostrovsky, R., Reyzin, L., Smith, A.: Fuzzy extractors: How to
  generate strong keys from biometrics and other noisy data.
\newblock SIAM journal on computing \textbf{38}(1), 97--139 (2008)

\bibitem{dworkin2015sha3}
Dworkin, M.J.: {SHA-3 Standard: Permutation-based Hash and Extendable-output
  Functions}.
\newblock NIST Pubs  (2015).
\newblock \doi{10.6028/NIST.FIPS.202}

\bibitem{keepass}
{EVERIS-NTT DATA Company}: { KeePass Code Review Results Report}.
\newblock Avilable at
  \url{https://joinup.ec.europa.eu/collection/eu-fossa-2/project-deliveries}
  (2016)

\bibitem{gringoli2019free}
Gringoli, F., Schulz, M., Link, J., Hollick, M.: {Free Your CSI: A Channel
  State Information Extraction Platform For Modern Wi-Fi Chipsets}.
\newblock In: Proceedings of the 13th International Workshop on Wireless
  Network Testbeds, Experimental Evaluation \& Characterization (2019)

\bibitem{halperin2011tool}
Halperin, D., Hu, W., Sheth, A., Wetherall, D.: {Tool release: Gathering 802.11
  n traces with channel state information}.
\newblock ACM SIGCOMM Computer Communication Review \textbf{41}(1), 53--53
  (2011)

\bibitem{RFC8110}
Harkins, D., Kumari, W.: {Opportunistic Wireless Encryption}.
\newblock RFC 8110 (2017).
\newblock \doi{10.17487/RFC8110}.
\newblock \urlprefix\url{https://www.rfc-editor.org/rfc/rfc8110.html}

\bibitem{harmon2008implementation}
Harmon, K., Johnson, S., Reyzin, L.: {An implementation of syndrome encoding
  and decoding for binary BCH codes, secure sketches and fuzzy extractors}.
\newblock Available at \url{https://www.cs.bu.edu/~reyzin/code/fuzzy.html}
  (2008)

\bibitem{hocquenghem1959codes}
Hocquenghem, A.: Codes correcteurs d’erreurs.
\newblock Chiffres \textbf{2}(2), 147--56 (1959)

\bibitem{80211_2016}
{IEEE}: {IEEE Std 802.11-2016 Part 11: Wireless LAN Medium Access Control (MAC)
  and Physical Layer (PHY) Specifications}  (2016).
\newblock \doi{10.1109/IEEESTD.2016.7786995}

\bibitem{interpol-covid}
INTERPOL: {COVID-19 Cybercrime Analysis Report}.
\newblock Available at
  \url{https://www.interpol.int/en/News-and-Events/News/2020/INTERPOL-report-shows-alarming-rate-of-cyberattacks-during-COVID-19}
  (2020)

\bibitem{jung2005time}
Jung, P., Wunder, G.: On time-variant distortions in multicarrier transmission
  with application to frequency offsets and phase noise.
\newblock IEEE Transactions on Communications \textbf{53}(9), 1561--1570 (2005)

\bibitem{kaliski2000rfc2898}
Kaliski, B.: {PKCS \#5: Password-based cryptography specification version 2.0}.
\newblock RFC 2898 (2000).
\newblock \doi{10.17487/RFC2898}.
\newblock \urlprefix\url{https://www.rfc-editor.org/rfc/rfc2898.html}

\bibitem{keyfactor}
{Kilgallin}, J., {Vasko}, R.: {Factoring RSA Keys in the IoT Era}.
\newblock In: IEEE International Conference on Trust, Privacy and Security in
  Intelligent Systems and Applications (2019)

\bibitem{rfc3526}
Kivinen, T., Kojo, M.: {More Modular Exponential (MODP) Diffie-Hellman groups
  for Internet Key Exchange (IKE)}.
\newblock RFC 3526 (2003).
\newblock \doi{10.17487/RFC3526}.
\newblock \urlprefix\url{https://www.rfc-editor.org/rfc/rfc3526.html}

\bibitem{kotaru2015spotfi}
Kotaru, M., Joshi, K., Bharadia, D., Katti, S.: {SpotFi: Decimeter Level
  Localization Using WiFi}.
\newblock In: Proceedings of the 2015 ACM Conference on Special Interest Group
  on Data Communication, pp. 269--282 (2015)

\bibitem{maurer1993secret}
Maurer, U.M.: Secret key agreement by public discussion from common
  information.
\newblock {IEEE Transactions on Information Theory}  (1993)

\bibitem{leastsquare}
Meyer, F.: {A single header-only C++ library for least squares fitting}.
\newblock Available at \url{https://github.com/Rookfighter/least-squares-cpp}
  (2019)

\bibitem{rappaport2001wireless}
Rappaport, T.: {Wireless Communications: Principles and Practice} pp. 165--166
  (2001)

\bibitem{autokey}
Reaz, K., Wunder, G.: {Wireless Channel-based Autonomous Key Management for IoT
  (AutoKEY) on WiSHFUL Testbed}.
\newblock Avilable at
  \url{http://www.wishful-project.eu/sites/default/files/AutoKEY-leaflet.pdf}
  (2017)

\bibitem{shannon1949communication}
Shannon, C.E.: {Communication theory of secrecy systems}.
\newblock {The Bell System Technical Journal} \textbf{28}(4), 656--715 (1949)

\bibitem{thai2018secret}
Thai, C.D.T., Lee, J., Prakash, J., Quek, T.Q.: {Secret Group-Key Generation at
  Physical Layer for Multi-Antenna Mesh Topology}.
\newblock IEEE Trans. on Information Forensics and Security  (2018)

\bibitem{KRACK}
Vanhoef, M., Piessens, F.: {Key Reinstallation Attacks: Forcing Nonce Reuse in
  WPA2}.
\newblock In: Proceedings of the 2017 ACM SIGSAC Conference on Computer and
  Communications Security. ACM (2017)

\bibitem{vasisht2016decimeter}
Vasisht, D., Kumar, S., Katabi, D.: {Decimeter-Level Localization with a Single
  WiFi Access Point}.
\newblock In: 13th USENIX Symposium on Networked Systems Design and
  Implementation (NSDI 16), pp. 165--178 (2016)

\bibitem{wpsCert}
Vieböck, S.: {Wi-Fi Protected Setup (WPS) PIN brute force vulnerability}.
\newblock CERT Vulnerability Note VU 723755.
\newblock \urlprefix\url{{https://www.kb.cert.org/vuls/id/723755/}}

\bibitem{wang2012cooperative}
Wang, Q., Xu, K., Ren, K.: {Cooperative Secret Key Generation from Phase
  Estimation in Narrowband Fading Channels}.
\newblock IEEE Journal on selected areas in communications \textbf{30}(9),
  1666--1674 (2012)

\bibitem{wheeler2016zxcvbn}
Wheeler, D.L.: {zxcvbn: Low-budget Password Strength Estimation}.
\newblock In: {25th USENIX Security Symposium}, pp. 157--173 (2016)

\bibitem{EasyConnect}
{Wi-Fi Alliance}: {Wi-Fi Easy Connect}.
\newblock Available at
  \url{https://www.wi-fi.org/discover-wi-fi/wi-fi-easy-connect}, visited on
  (23/10/2019)

\bibitem{OWESpecs}
{Wi-Fi Alliance}: {Opportunistic Wireless Encryption Specification}.
\newblock Specification v1.0 (2019)

\bibitem{wps}
{Wi-Fi Alliance}: {Wi-Fi Protected Setup Version 2.0.2} (2020)

\bibitem{alliance2020wpa3}
{Wi-Fi Alliance}: {WPA3 Specification Version 3.0} (2020)

\bibitem{wu2015phaseu}
Wu, C., Yang, Z., Zhou, Z., Qian, K., Liu, Y., Liu, M.: {PhaseU: Real-time LOS
  identification with WiFi}.
\newblock In: IEEE conference on computer communications, pp. 2038--2046. IEEE
  (2015)

\bibitem{wyner1975wiretap}
Wyner, A.D.: {The Wire-Tap Channel}.
\newblock {The Bell System Technical Journal} \textbf{54}(8) (1975)

\bibitem{xi2016instant}
Xi, W., Qian, C., Han, J., Zhao, K., Zhong, S., Li, X.Y., Zhao, J.: {Instant
  and Robust Authentication and Key Agreement among Mobile Devices}.
\newblock In: Proceedings of the 2016 ACM SIGSAC Conference on Computer and
  Communications Security (2016)

\bibitem{xie2016atheros}
Xie, Y., Li, Z., Li, M.: {Precise Power Delay Profiling with Commodity WiFi}.
\newblock MobiCom '15. ACM (2015).
\newblock \doi{10.1145/2789168.2790124}

\bibitem{zenger2015security}
Zenger, C., Zimmer, J., Paar, C.: Security analysis of quantization schemes for
  channel-based key extraction.
\newblock In: proceedings of the 12th EAI International Conference on Mobile
  and Ubiquitous Systems: Computing, Networking and Services (2015)

\bibitem{zenger2014novel}
Zenger, C.T., Chur, M.J., Posielek, J.F., Paar, C., Wunder, G.: A novel key
  generating architecture for wireless low-resource devices.
\newblock In: 2014 International Workshop on Secure Internet of Things. IEEE
  (2014)

\bibitem{zhu2018pi}
Zhu, H., Zhuo, Y., Liu, Q., Chang, S.: {$\pi$-splicer: Perceiving accurate CSI
  phases with commodity WiFi devices}.
\newblock IEEE Transactions on Mobile Computing \textbf{17}(9), 2155--2165
  (2018)

\bibitem{zinoviev1996solution}
Zinoviev, V.: {On the solution of equations of degree $ \leq 10$ over finite
  fields $GF (2^m)$}.
\newblock Rapports de recherche- INRIA  (1996)

\end{thebibliography}
\end{document}